\documentclass[aps,prl,twocolumn,floatfix]{revtex4-1}

\usepackage{graphicx}
\usepackage{amssymb,amsfonts,amsmath}
\usepackage{epsf}
\usepackage{subfigure}
\usepackage{epstopdf}
\usepackage{mathrsfs}

\newcommand{\be}{\begin{equation}}
\newcommand{\ee}{\end{equation}}
\newcommand{\bea}{\begin{eqnarray}}
\newcommand{\eea}{\end{eqnarray}}

\begin{document}

\title{\bf Mechanochemical fluctuation theorem and thermodynamics of self-phoretic motors}

\author{Pierre Gaspard$^1$ and Raymond Kapral$^2$}
\email{gaspard@ulb.ac.be,\,rkapral@chem.utoronto.ca }
%\email{rkapral@chem.utoronto.ca}
\affiliation{$^1$Center for Nonlinear Phenomena and Complex Systems,
Universit{\'e} Libre de Bruxelles, Code Postal 231, Campus Plaine, B-1050 Brussels, Belgium \\
$^2$ Chemical Physics Theory Group, Department of Chemistry, University of Toronto, Toronto, Ontario M5S 3H6, Canada}

\begin{abstract}
Microscopic dynamical aspects of the propulsion of nanomotors by self-phoretic mechanisms are considered. Propulsion by self-diffusiophoresis relies on the mechanochemical coupling between the fluid velocity field and the concentration fields induced by asymmetric catalytic reactions on the motor surface. The consistency between the thermodynamics of this coupling and the microscopic reversibility of the underlying molecular dynamics is investigated. For this purpose coupled Langevin equations for the translational, rotational, and chemical fluctuations of self-phoretic motors are derived. A mechanochemical fluctuation theorem for the joint probability to find the motor at position ${\bf r}$ after $n$ reactive events have occurred during the time interval $t$ is also derived. An important result that follows from this analysis is the identification of an effect that is reciprocal to self-propulsion by diffusiophoresis, which leads to the possibility of fuel synthesis by mechanochemical coupling to external force and torque.
\end{abstract}

\maketitle

Recently, synthetic micromotors powered by different self-phoretic mechanisms have been constructed and studied experimentally \cite{PKOSACMLC04,FBAMO05,W13,WDAMS13,SSK15}. Self-propulsion is achieved by the generation of local gradients of chemical concentrations, electrochemical potential, or temperature, which produce the force driving the motor \cite{A89,GLA05,RK07,TK09,K13,CRRK14}.  This is the case in particular for Janus motors with catalytic and chemically-inactive hemispheres, moving by diffusiophoresis in a solution with out-of-equilibrium concentrations of fuel and product \cite{CRRK14,SS12,dBK13,HSK16}.  The propulsion mechanism is based on the mechanochemical coupling between the fluid velocity around the motor and the concentration fields induced by the reaction taking place on the catalytic hemisphere. A fundamental issue that arises in this context is the consistency between the thermodynamics of this coupling and the microreversibility of the underlying molecular dynamics.  The challenge is that the synthetic motors have micro- or nanometric sizes and, therefore, are subjected to thermal fluctuations due to the atomic structure of matter.

In this letter, we address this issue by deducing coupled Langevin equations for the translational, rotational, and chemical fluctuations of self-phoretic motors, along with a mechanochemical fluctuation theorem.  Since the fluctuation theorem is a consequence of microreversibility, we can identify the effect that is reciprocal to the self-diffusiophoretic propulsion.  In this way, we show that the chemical reaction can be reversed and the synthesis of fuel from product can be achieved by applying an external force while controlling the directionality of the Janus particle.  This reciprocal effect is  analogous to what is observed at the nanoscale in molecular motors \cite{JAP97,ITANYYK04,GG07,LM09}.

With this aim in mind, through a systematic analysis, we have determined the boundary conditions coupling the fluid velocity ${\bf v}=(v_x,v_y,v_z)$ and the concentration fields $c_k(x,y,z)$ of the different chemical species $k$ on the particle surface within the thin-layer approximation.  The molecules of species $k$ in solution are assumed to interact with the surface of the Janus motor by the potential energy $u_k(x,y,z)$ that vanishes beyond the range $\delta$, which is assumed to be small with respect to the particle radius $R$.  In this case, the approximation of a locally flat surface holds and the coupled Stokes and diffusion equations can be solved in the thin interaction layer in order to obtain the effective boundary conditions that the fields would satisfy if the layer were arbitrarily thin.  Previous work on the thin-layer approximation has been devoted to diffusiophoresis in concentration gradients applied at the macroscale, larger than the particle size \cite{A89}.  For self-diffusiophoresis, the boundary conditions on the concentration fields are modified by the reaction at the surface of the moving particle.  Taking the $z$-direction perpendicular and the $(x,y)$-directions parallel to the surface, the slip velocity and the diffusive fluxes are given for such modified conditions by
\bea
&& v_\alpha\vert_{z=0} = \left( b \, \partial_z v_\alpha - \sum_k b_k \partial_\alpha c_k\right)_{z=0}  \quad(\alpha=x,y) \, ,\label{vx-bc}\\
&& -D_k \partial_z c_k\vert_{z=0} = \nu_k \, w(x,y) - \sum_{\alpha=x,y} \partial_{\alpha}\left(\frac{\eta\, b_k}{k_{\rm B}T\, b}\, c_k \, v_{\alpha}\right)_{z=0}  \label{ck-bc}
\eea
up to corrections of higher powers in the thickness $\delta$ of the layer and in the gradient $\partial_{\alpha}$ parallel to the surface.  In Eqs.~(\ref{vx-bc}) and~(\ref{ck-bc}), $b$ is the slip length \cite{AB06},
\be
b_k = \frac{k_{\rm B}T}{\eta} \left( K_k^{(1)} + b \, K_k^{(0)} \right)
\label{b_k}
\ee
with
\be K_k^{(n)} \equiv \int_0^{\delta} dz \, z^n \, \left[ {\rm e}^{-\beta u_k(z)}-1\right]
\label{K_k}
\ee
is the coefficient of coupling of the surface concentration gradient $\partial_xc_k$ to the slip velocity \cite{A89}, $\eta$ the fluid shear viscosity, $T$ the temperature, $k_{\rm B}$ Boltzmann's constant, $D_k$ the diffusion coefficient of solute $k$, $w(x,y)$ the surface reaction rate, and $\nu_k$ the stoichiometric coefficient of species $k$ in the reaction.  The last term of Eq.~(\ref{ck-bc}) is responsible for the reciprocal effect of the fluid velocity back onto the reaction rate at this level of description.

 Employing the boundary conditions~(\ref{vx-bc})-(\ref{ck-bc}) at the surface of the Janus motor, and using Faxen's theorem in conjunction with a fluctuating hydrodynamics formulation \cite{BM74,ABM75}, the following Langevin equation is deduced for a spherical particle:
\be
m\frac{d{\bf v}}{dt} = -\gamma\, {\bf v}  + {\bf F}_{\rm d} + {\bf F}_{\rm ext} + {\bf F}_{\rm fl}(t) \, ,
\label{Langevin-eq}
\ee
where $m$ is the mass of the Janus motor, ${\bf v}=d{\bf r}/dt$ its velocity, $\gamma = 6\pi\eta R \,  ({1+\frac{2b}{R}})/({1+\frac{3b}{R}})$
the translational friction coefficient related by Einstein's relation $\gamma=(\beta D)^{-1}$ to the diffusion coefficient $D$ \cite{ABM75},
\be
{\bf F}_{\rm d} = \frac{6\pi\eta R}{1+\frac{3b}{R}}  \sum_k b_k  \overline{({\boldsymbol{\mathsf 1}}-{\bf n}{\bf n})\cdot\pmb{\nabla}c_k({\bf r},t)}^{\rm s}
\ee
the diffusiophoretic force involving the surface average $\overline{(\cdot)}^{\rm s} =\int_{r=R}(\cdot)dS/(4\pi R^2)$ ($\bf n$ being a unit vector perpendicular to the surface), ${\bf F}_{\rm ext}$ an external force (e.g. the gravitational force) \cite{TK09}, and ${\bf F}_{\rm fl}(t)$ the Langevin fluctuating force.  The diffusiophoretic force is directed along the axis of the Janus motor, specified by the unit vector~$\bf u$: ${\bf F}_{\rm d}=F_{\rm d}{\bf u}$.  Moreover, this force is  proportional to the mean reaction rate $W_{\rm rxn}$ through the surface gradient of the concentration fields.  Accordingly, we introduce the diffusiophoretic coupling coefficient $\chi\equiv F_{\rm d}/(\gamma W_{\rm rxn})$.  Using the definitions of ${\bf F}_{\rm d}$ and $\gamma$ given above, the explicit dependence of $\chi$ on the slip length $b$ can be written in the form $\chi\sim(C^{(1)}+b C^{(0)})/(1+\frac{2b}{R})$ where the quantities $C^{(n)}$ are given in terms of the constants~(\ref{K_k}) and the molecular diffusivities~$D_k$ of species $k={\rm A},{\rm B}$. From this expression we see that $\chi$ has a well-defined value in both the limit $b=0$ for stick boundary conditions and the limit $b=\infty$ for perfect slip boundary conditions.  An enhancement of the diffusiophoretic effects is expected if the hydrophobicity is large because $\chi\sim (\delta/R)^\ell$ with $\ell=2$ if $b=0$, but with $\ell=1$ if $b\to\infty$.

In the overdamped limit, the Langevin equation~(\ref{Langevin-eq}) becomes
\be
\frac{d{\bf r}}{dt} = V_{\rm d} \, {\bf u} + \beta D\, {\bf F}_{\rm ext} + {\bf V}_{\rm fl}(t) \, . \label{eq-transl}
\ee
where ${\bf r}=(x,y,z)$ is the particle position, $V_{\rm d}=\chi W_{\rm rxn}=F_{\rm d}/\gamma$ is the diffusiophoretic velocity and ${\bf V}_{\rm fl}(t) = {\bf F}_{\rm fl}(t)/\gamma$  the fluctuating velocity.

The orientation $\bf u$ of the Janus particle is ruled by the following rotational overdamped Langevin equation:
\be
\frac{d{\bf u}}{dt} = -\frac{1}{\gamma_{\rm rot}} \, {\bf u}\times \left[{\bf T}_{\rm ext} + {\bf T}_{\rm fl}(t) \right]  \, , \label{eq-rot}
\ee
where $\gamma_{\rm rot} = {8\pi\eta R^3}/({1+\frac{3b}{R}})$
is the rotational friction coefficient \cite{F76}, ${\bf T}_{\rm ext} = \mu{\bf u}\times{\bf B}$ is an external torque due to an external magnetic field $\bf B$ exerted on a magnetic dipole $\mu{\bf u}$ attached to the particle or the gravitational field acting on the nonuniform mass density of the Janus particle \cite{CE13}, and ${\bf T}_{\rm fl}(t)$ is the Langevin fluctuating torque associated with the rotational diffusion coefficient $D_{\rm rot}=k_{\rm B}T/\gamma_{\rm rot}$.  Since the Janus motor is assumed to be spherical, there is no torque due to diffusiophoresis.  We note that the external force and torque derive from the potential energy $U({\bf r},{\bf u})=-{\bf F}_{\rm ext}\cdot{\bf r}-\mu{\bf B}\cdot{\bf u}$.

In order to describe the mechanochemical coupling, Eqs.~(\ref{eq-transl}) and~(\ref{eq-rot}) must be supplemented by a stochastic equation for the chemical reaction.  Here, we consider the simple reaction ${\rm A}\rightleftharpoons{\rm B}$, where A is the fuel and B the product, so that the mean reaction rate is given by $W_{\rm rxn}=\Gamma (k_+\bar{c}_{\rm A}-k_-\bar{c}_{\rm B})$ in terms of the rate constants $k_{\pm}$ and the concentrations $\bar{c}_{\rm A}$ and $\bar{c}_{\rm B}$ at an arbitrarily large distance from the Janus particle, up to a dimensionless constant $\Gamma$.  The mean reaction rate vanishes at chemical equilibrium when $\bar{c}_{\rm A}/\bar{c}_{\rm B}=k_-/k_+$.  In order to satisfy microreversibility, the chemical stochastic equation must take the form,
\be
\frac{dn}{dt} = W_{\rm rxn} + \beta\chi D_{\rm rxn} {\bf u}\cdot{\bf F}_{\rm ext} + W_{\rm fl}(t) \, ,
\label{eq-rxn}
\ee
where the second term on the right ($\beta\chi D_{\rm rxn} {\bf u}\cdot{\bf F}_{\rm ext}$) is a reciprocal contribution of the external force back onto the reaction rate due to the diffusiophoretic coupling and proportional to the reaction diffusivity $D_{\rm rxn}$.  The velocity and rate fluctuations are coupled Gaussian white noises characterized by
\bea
&&\langle {\bf V}_{\rm fl}(t)\rangle = 0 \ , \qquad \langle W_{\rm fl}(t)\rangle = 0 \, , \\
&&\langle {\bf V}_{\rm fl}(t)\otimes {\bf V}_{\rm fl}(t')\rangle = 2D \, \delta(t-t') \, {\boldsymbol{\mathsf 1}} \, , \\
&&\langle W_{\rm fl}(t)\, W_{\rm fl}(t')\rangle = 2D_{\rm rxn} \, \delta(t-t') \, , \\
&&\langle {\bf V}_{\rm fl}(t)\, W_{\rm fl}(t')\rangle = 2\chi D_{\rm rxn} \, {\bf u} \, \delta(t-t') \, ,
\eea
where $\otimes$ denotes the tensorial product and ${\boldsymbol{\mathsf 1}}$ the $3\times 3$ identity matrix.
The necessity of including the reciprocal contribution can be seen by considering the evolution equations for the mean position ${\bf r}$ and number $n$. Letting ${\bf X}=({\bf r},\; n)$, these equations are
\begin{equation}
\frac{d\langle{\bf X}\rangle}{dt} = {\boldsymbol{\mathsf L}}\cdot{\bf A},
\end{equation}
where ${\bf A} = ({\bf A}_{\rm mech} , \, A_{\rm rxn}  )$ is the vector of the generalized thermodynamic forces comprising the mechanical affinity, ${\bf A}_{\rm mech} = \beta \, {\bf F}_{\rm ext}$, and chemical affinity, $A_{\rm rxn} = {W_{\rm rxn}}/{D_{\rm rxn}}$~\cite{P67,GM84}, while the matrix ${\boldsymbol{\mathsf L}}$ is given by
\be
{\boldsymbol{\mathsf L}} =
\left(
\begin{array}{cc}
D \, {\boldsymbol{\mathsf 1}} & \chi\, D_{\rm rxn} \, {\bf u} \\
\chi\, D_{\rm rxn} \, {\bf u} & D_{\rm rxn}
\end{array}
\right),
\label{L}
\ee
with ${\boldsymbol{\mathsf L}}={\boldsymbol{\mathsf L}}^{\rm T}$ to be consistent with Onsager's reciprocal relations.
In order to satisfy the second law of thermodynamics, the diffusivities should satisfy $D\ge 0$, $D_{\rm rxn} \ge 0$, and $D\ge \chi^2 D_{\rm rxn}$.
The control parameters are the mean reaction rate $W_{\rm rxn}$ determined by the solute concentrations, the external force ${\bf F}_{\rm ext}$, and the external torque~${\bf T}_{\rm ext}$.  An important aspect is that only the mean reaction rate and the external force can drive the Janus particle into a nonequilibrium steady state.  Indeed, the external torque has here the sole effect of aligning the Janus particle parallel to the external magnetic or gravitational field, but does not generate a gyration of the particle as in Ref.~\cite{SJS11}.  Accordingly, the probability distribution of the particle orientation reaches equilibrium after the rotational relaxation time $\tau_{\rm rot}=1/(2D_{\rm rot})$ and no longer contributes to the entropy production rate,
\be
\frac{1}{k_{\rm B}}\frac{d_{\rm i}S}{dt} =\beta\,{\bf F}_{\rm ext}\cdot\langle{\bf\dot r}\rangle + A_{\rm rxn}\,\langle\dot n\rangle \geq 0 \,.
\label{entr-prod}
\ee

\begin{figure}[h!]
\centering
\includegraphics[width=\columnwidth]{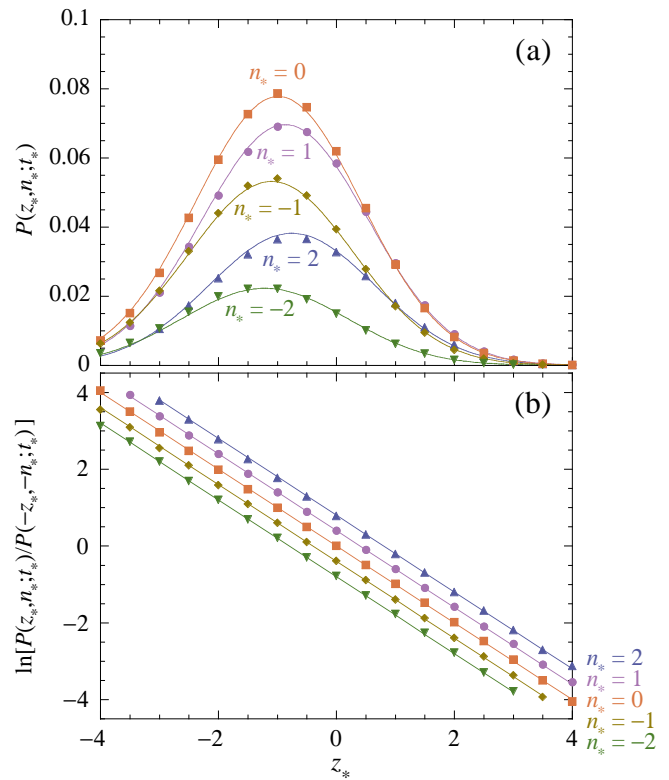}
\caption{Janus particle subjected to an external force and magnetic field oriented in the $z$-direction: (a) Probability density $P(z_*,n_*;t_*)$ with $n_*=n\sqrt{D_{\rm rot}/D_{\rm rxn}}=2,1,0,-1,-2$ versus the rescaled displacement $z_*=z\sqrt{D_{\rm rot}/D}$ at the rescaled time $t_*=D_{\rm rot}t=1$ for the parameter values $p=\beta\mu B=1$, $f=\beta F\sqrt{D/D_{\rm rot}}=-1$, $w=W_{\rm rxn}/\sqrt{D_{\rm rxn}D_{\rm rot}}=0.4$, and $c=\chi\sqrt{D_{\rm rxn}/D}=0.4$. (b) Verification of the mechanochemical fluctuation relation~(\ref{FT}) in the same conditions.  The probability ratio is calculated if $P(z_*,n_*;t_*)$ and $P(-z_*,-n_*;t_*)$ are larger than $10^{-4}$. The dots are the results of a numerical simulation with an ensemble of $10^{7}$ trajectories and an integration with the time step $dt_*=10^{-3}$, using the method described in Supplementary Material~\cite{SM}.  The lines depict the theoretical expectations.}\label{fig1}
\end{figure}

The mechanochemical fluctuation theorem corresponding to the entropy production~(\ref{entr-prod}) is given by
\be
\frac{P({\bf r},n;t)}{P(-{\bf r},-n;t)} \simeq \exp\left(\beta\, {\bf F}_{\rm ext}\cdot{\bf r} + A_{\rm rxn} \, n \right)
\label{FT}
\ee
for the joint probability density $P({\bf r},n;t)$ to find the motor at the position $\bf r$ after $n$ reactive events have occurred during the time interval $t$.  This latter should be longer than the rotational relaxation time, as well as the characteristic time of solute molecular diffusion (which is of the order of $\tau_{{\rm diff},k}\sim R^2/D_k$ in the diffusion-limited regime).  The fluctuation theorem~(\ref{FT}) can be deduced from the Fokker-Planck equation for the coupled Langevin equations by using methods of large-deviation theory \cite{LM09,LS99,G13}.  This theorem extends previous relations~\cite{SJS11,KSRS15,FPBCK16,PKS16} by including the chemical fluctuations, which are essential to obtain all of the contributions to the entropy production and prove its non-negativity~(\ref{entr-prod}) by Jensen's inequality $\langle\exp x\rangle\geq\exp\langle x\rangle$.  Figure~\ref{fig1} shows that the mechanochemical fluctuation theorem is satisfied.

Suppose that the particle is subjected to an external force in the $z$-direction ${\bf F}_{\rm ext}=(0,0,F)$, as well as to the external magnetic field ${\bf B}=(0,0,B)$ so that the particle is oriented on average in that direction: $\langle u_z\rangle={\rm coth}(\beta\mu B)-1/(\beta\mu B)$.  Often, only the position is observed while the rate is very large. Since the probability distribution becomes Gaussian after a long enough time by the central limit theorem, we recover the effective fluctuation relation \cite{FPBCK16} for the displacement along the $z$-direction
\be
\frac{{\mathscr P}(z;t)}{{\mathscr P}(-z;t)} \simeq \exp \frac{F_{\rm eff}z}{k_{\rm B}T_{\rm eff}} \, ,
\ee
which is expressed in terms of an effective force $F_{\rm eff}=F+F_{\rm d}\langle u_z\rangle$ resulting from the external and diffusiophoretic forces, and the effective temperature $T_{\rm eff}=T\left[1+(V_{\rm d}^2/D)\int_0^{\infty} C_{zz}(t) dt\right]$ where $C_{zz}(t)\equiv \langle[u_z(0)-\langle u_z\rangle][u_z(t)-\langle u_z\rangle]\rangle$ is the time-dependent correlation function of the orientation along the $z$-direction.  In the absence of an external force and torque ($F=0$ and $B=0$), we also recover the known result that diffusion is enhanced due to the self-phoretic effect, the effective translational diffusion coefficient being given by $D_{\rm eff}=D+V_{\rm d}^2/(6D_{\rm rot})$.

\begin{figure}[htbp]
\includegraphics[scale=0.55]{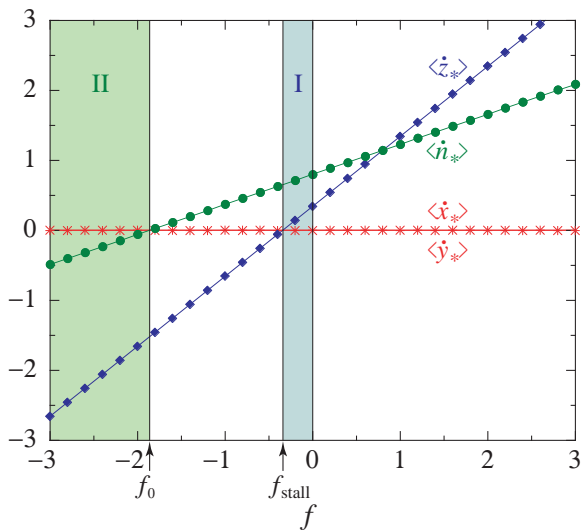}
\caption{Janus particle subjected to an external force and magnetic field oriented in the $z$-direction: The mean values of the fluctuating rescaled velocities ${\bf\dot r}_*={\bf\dot r}/\sqrt{DD_{\rm rot}}$ and rate $\dot n_*=\dot n/\sqrt{D_{\rm rxn}D_{\rm rot}}$ versus the rescaled magnitude of the external force $f=\beta F\sqrt{D/D_{\rm rot}}$ for the parameter values $p=\beta\mu B=2$, $w=W_{\rm rxn}/\sqrt{D_{\rm rxn}D_{\rm rot}}=0.8$, and $c=\chi\sqrt{D_{\rm rxn}/D}=0.8$.  The dots show the results of a numerical simulation with a statistics of $10^5$ trajectories integrated over the time interval $t_*=10$, using the method described in Supplementary Material~\cite{SM}.  $f_{\rm stall}$ denotes the rescaled stall force and $f_0$ the threshold between fuel synthesis and consumption.  The Janus particle is propelled against the external force in the interval I.  Fuel synthesis happens in the interval II.}
\label{fig2}
\end{figure}

In the presence of external force and torque, the Janus particle can move against the external force, as shown in Fig.~\ref{fig2}.  The condition is that the force $F$ takes a value between the stall force $F_{\rm stall}=-V_{\rm d}\langle u_z\rangle/(\beta D)$ and zero.

A key point is that the fluctuation theorem~(\ref{FT}) would not hold without the reciprocal term due to the diffusiophoretic coupling $\chi$ in Eq.~(\ref{eq-rxn}).  A most important consequence of this term is that the chemical reaction can be reversed if a large enough external force is exerted in a direction opposite to self-propulsion: $F< F_0=-W_{\rm rxn}/(\beta\chi D_{\rm rxn}\langle u_z\rangle)$.  In this regime, fuel is synthesized from product.  The thermodynamic efficiency of synthesis $\eta_{\rm c}\equiv-A_{\rm rxn}\langle \dot n\rangle/(\beta F \langle\dot z\rangle)$ \cite{JAP97} can reach the maximum value $\eta_{\rm c}^{\rm (max)}=(1-\sqrt{1-q^2})/(1+\sqrt{1-q^2})=0.25\, q^2+O(q^4)$ with $-1\leq q\equiv\chi\langle u_z\rangle\sqrt{D_{\rm rxn}/D}\leq +1$.  Therefore, the larger the diffusiophoretic coupling coefficient $\chi$, the larger the efficiency.  Applying a counter force of sufficient magnitude to a motor oriented by a torque should yield the conversion of product to fuel, which could be verified experimentally. A corollary of this result is that the action of diffusiophoretic micropumps \cite{SPOASDCMS14} can also be reversed and fuel synthesized if a large enough pressure is applied to a product solution flowing through a pore with part of its inner surface coated by catalyst.  The possibility of fuel synthesis by the mechanochemical coupling is the reciprocal effect of self-propulsion (or pumping) and constitutes a principal prediction of this Letter.

The previous results can be generalized to other self-electrophoretic or self-thermophoretic motors, as well as to non-spherical shapes where extra couplings are expected between translation, rotation, and reaction.

%%%%%%%%%%%%%%%%%%%%%%%%%%%%%%%%%%%%%%%%%%%%%%%%%%%%%%%%%%%
%\section*{Acknowledgments}
The authors thank P. Grosfils and M.-J. Huang for stimulating discussions.  Financial support from the International Solvay Institutes for Physics and Chemistry, the Universit\'e libre de Bruxelles (ULB), the Fonds de la Recherche Scientifique~-~FNRS under the Grant PDR~T.0094.16 for the project ``SYMSTATPHYS", the Belgian Federal Government under the Interuniversity Attraction Pole project P7/18 ``DYGEST", and the Natural Sciences and Research Council of Canada is acknowledged.

\vskip 10pt
%%%%%%%%%%%%%%%%%%%%%%%%%%%%%%%%%%%%%%%%%%%%%%%%%

%%%%%%%%%%%%%%%%%%%%%%%%%%%%%%%%%%%%%%%%%%%%%%%%%%%%%%%%%%%
\begin{widetext}

\vskip 0.5 cm

\appendix{\centerline{\large\sf Supplementary Material}}

\vskip 0.3 cm

The aim of this supplementary material is to describe the numerical method used to simulate the process based on the coupled overdamped Langevin equations given by
\bea
\frac{d{\bf u}}{dt} &=& \pmb{\omega}_{\rm ext}\times{\bf u} +  \pmb{\omega}_{\rm fl}(t)\times{\bf u} \, , \\
\frac{d{\bf r}}{dt} &=& \chi \, W_{\rm rxn} \, {\bf u} + \beta D\, {\bf F}_{\rm ext} + {\bf V}_{\rm fl}(t) \, , \\
\frac{dn}{dt} &=& W_{\rm rxn} + \beta\chi D_{\rm rxn} {\bf u}\cdot{\bf F}_{\rm ext} + W_{\rm fl}(t) \, ,
\eea
with the angular velocity due to the external magnetic field $\bf B$ exerted on the magnetic dipole $\mu$ of the Janus particle
\be
\pmb{\omega}_{\rm ext} = \frac{1}{\gamma_{\rm rot}}\, {\bf T}_{\rm ext}  = \frac{\mu}{\gamma_{\rm rot}}\, {\bf u}\times{\bf B} \, .
\ee
The fluctuating angular velocity $\pmb{\omega}_{\rm fl}(t)={\bf T}_{\rm fl}(t)/\gamma_{\rm rot}$, the fluctuating velocity ${\bf V}_{\rm fl}(t)={\bf F}_{\rm fl}(t)/\gamma$, and the fluctuating reaction rate $W_{\rm fl}(t)$ are coupled Gaussian white noises characterized by
\bea
&&\langle \pmb{\omega}_{\rm fl}(t)\rangle = 0 \ , \qquad \langle {\bf V}_{\rm fl}(t)\rangle = 0 \ , \qquad \langle W_{\rm fl}(t)\rangle = 0 \, , \\
&&\langle \pmb{\omega}_{\rm fl}(t)\otimes \pmb{\omega}_{\rm fl}(t')\rangle = 2D_{\rm rot} \, \delta(t-t') \, {\boldsymbol{\mathsf 1}} \, , \\
&&\langle {\bf V}_{\rm fl}(t)\otimes {\bf V}_{\rm fl}(t')\rangle = 2D \, \delta(t-t') \, {\boldsymbol{\mathsf 1}} \, , \\
&&\langle W_{\rm fl}(t)\, W_{\rm fl}(t')\rangle = 2D_{\rm rxn} \, \delta(t-t') \, , \\
&&\langle {\bf V}_{\rm fl}(t)\, W_{\rm fl}(t')\rangle = 2\chi D_{\rm rxn} \, {\bf u} \, \delta(t-t') \, .
\eea

We notice that the stochastic equation for rotation is decoupled from the ones for translation and rotation.  The fluctuating rotational motion is simulated by the method of quaternions \cite{IBdO15,DBD15}.  The different noises can thus be decoupled as follows:
\bea
\pmb{\omega}_{\rm fl}(t) &=& \sqrt{2D_{\rm rot}}\, \pmb{\xi}_{\rm rot}(t) \, , \\
{\bf V}_{\rm fl}(t) &=& \sqrt{2D}\, \xi_1(t) \, {\bf u}_1 + \sqrt{2D}\, \xi_2(t) \, {\bf u}_2 +\left[\quad\ \ \, \sqrt{D\lambda_+}\, \xi_3(t) -\ \ \varsigma\sqrt{D\lambda_-}\, \xi_4(t)\right] \, {\bf u}_3 \, , \\
W_{\rm fl}(t) &=& \qquad\qquad\qquad\qquad\qquad\qquad\qquad\ \varsigma\sqrt{D_{\rm rxn}\lambda_+}\, \xi_3(t) + \sqrt{D_{\rm rxn}\lambda_-}\, \xi_4(t) \, ,
\eea
in terms of the independent Gaussian white noises satisfying
\bea
&&\langle\pmb{\xi}_{\rm rot}(t)\rangle = 0 \, , \qquad \langle\pmb{\xi}_{\rm rot}(t)\otimes\pmb{\xi}_{\rm rot}(t')\rangle = \delta(t-t')\, {\boldsymbol{\mathsf 1}} \, , \qquad \langle\xi_i(t)\, \pmb{\xi}_{\rm rot}(t')\rangle = 0 \, , \label{xi-rot}\\
&&\langle\xi_i(t)\rangle = 0 \, , \qquad\ \  \langle\xi_i(t)\, \xi_j(t')\rangle = \delta(t-t') \, \delta_{ij} \, , \label{xi-TR}
\eea
the parameters
\be
\lambda_{\pm}=1\pm\sqrt{\chi^2D_{\rm rxn}/D} \, , \qquad \varsigma=\chi/\vert\chi\vert \, ,
\ee
and the unit vectors $\{{\bf u}_1,{\bf u}_2,{\bf u}_3={\bf u}\}$ attached to the frame of the Janus particle.  The unit vector ${\bf u}_3={\bf u}$ is oriented along the particle axis pointing from the inactive towards the catalytic hemisphere, while the unit vectors ${\bf u}_1$ and ${\bf u}_2$ are perpendicular to the axis.

Introducing the rescaled dimensionless variables
\bea
t_* &\equiv&D_{\rm rot}\, t \, , \label{t_*}\\
{\bf r}_* &\equiv&\sqrt{\frac{D_{\rm rot}}{D}}\, {\bf r} \, , \\
n_* &\equiv&\sqrt{\frac{D_{\rm rot}}{D_{\rm rxn}}}\, n \, ,
\eea
and the dimensionless parameters
\bea
p &\equiv& \beta\mu B \, , \\
f &\equiv& \beta F \sqrt{\frac{D}{D_{\rm rot}}}\, , \\
w &\equiv& \frac{W_{\rm rxn}}{\sqrt{D_{\rm rxn}D_{\rm rot}}}\, , \\
c &\equiv& \chi \sqrt{\frac{D_{\rm rxn}}{D}}\, ,
\eea
the stochastic differential equations read
\bea
\frac{d{\bf u}}{dt_*} &=& p\left({\bf u}\times{\bf 1}_{p}\right)\times{\bf u} +  \sqrt{2}\ \tilde{\pmb{\xi}}_{{\rm rot}}(t_*)\times{\bf u} \, , \\
\frac{d{\bf r}_*}{dt_*} &=& f\, {\bf 1}_{f}+ cw{\bf u} +\sqrt{2}\,\tilde\xi_{1}(t_*) \, {\bf u}_1 + \sqrt{2}\, \tilde\xi_{2}(t_*) \, {\bf u}_2 +\left[\sqrt{\lambda_+}\, \tilde\xi_{3}(t_*) -\varsigma\sqrt{\lambda_-}\, \tilde\xi_{4}(t_*)\right] \, {\bf u} \, , \\
\frac{dn_*}{dt_*} &=& w + c f {\bf u}\cdot{\bf 1}_{f} + \varsigma\sqrt{\lambda_+}\, \tilde\xi_{3}(t_*) + \sqrt{\lambda_-}\, \tilde\xi_{4}(t_*) \, ,
\eea
with the unit vectors ${\bf u}={\bf u}_3$, ${\bf 1}_p={\bf B}/\Vert{\bf B}\Vert$, ${\bf 1}_f={\bf F}_{\rm ext}/\Vert{\bf F}_{\rm ext}\Vert$, and the dimensionless independent Gaussian white noises $\tilde{\pmb{\xi}}_{\rm rot}(t_*)$ and $\tilde\xi_i(t_*)$ satisfying Eqs.~(\ref{xi-rot})-(\ref{xi-TR}) but with the time $t$ replaced by the dimensionless time $t_*$ defined by Eq.~(\ref{t_*}).

\vskip 0.3 cm

%%%%%%%%%%%%%%%%%%%%%%%%%%%%%%%%%%%%%%%%%%%%
\end{widetext}

\end{document}